\documentclass[useAMS,usenatbib]{mn2e}
\usepackage{epsfig}

\newif\ifAMStwofonts

\newcommand{\uc}{ULTRACAM\,}
\newcommand{\fast}{``fast''}
\newcommand{\slow}{``slow''}
\newcommand{\sloanu}{$u'$}
\newcommand{\sloang}{$g'$\,} 
\newcommand{\sloani}{$i'$\,}
\newcommand{\target}{V404\,Cyg}
\newcommand{\rtr}{$r_{\bf tr}$}
\newcommand{\rout}{$r_{\bf out}$}
\newcommand{\rsch}{$r_{\bf sch}$}
\newcommand{\Msun}{\,$\rm M_{\odot}$}
\newcommand{\Rsun}{\,$\rm R_{\odot}$}

\newcommand{\Rdisc}{\,$R_{\rm disc}$\,}
\newcommand{\kms}{\,$\rm km\,s^{-1}$}

\title[Multicolour observations of V404\,Cyg with \uc] 
{Multicolour observations of V404\,Cyg with \uc}

\author[T.\,Shahbaz et al.]
       {T.\,Shahbaz,$^{1}$\thanks{E-mail: tsh@ll.iac.es}
	V.S.\,Dhillon$^2$,
	T.R.\,Marsh$^3$,
	C.\,Zurita$^1$,
	C.\,A.\,Haswell$^4$,
	P.A.\,Charles$^3$ 
	\newauthor 
        R.I.\,Hynes$^5$,
        J.\,Casares$^1$ \\
$^1$Instituto de Astrof\'\i{}sica de Canarias, 38200 La Laguna,
    Tenerife, Spain \\
$^2$Department of Physics and Astronomy, University of Sheffield, 
    Sheffield, S3 7RH, UK  \\
$^3$Department of Physics and Astronomy, University of Southampton, 
    Southampton, SO17 1BJ, UK \\
$^4$Department of Physics and Astronomy, The Open University, Walton
    Hall, Milton Keynes, MK7 6AA \\
$^5$Astronomy Department, The University of Texas at Austin, 1
       University Station C1400, Austin, Texas 78712-0259, USA }
\pagerange{\pageref{firstpage}--\pageref{lastpage}}
\pubyear{2003}

\begin{document}
\maketitle
\begin{abstract}
\noindent

We present high time-resolution multicolour observations of the  quiescent soft
X-ray transient \target\ obtained with \uc.  Superimposed on the secondary
star's ellipsoidal modulation are large flares on timescales of a few hours, as
well as several distinct rapid flares on timescales of tens of mins. The rapid
flares, most of which show  further variability and unresolved peaks,   cover
shorter timescales than those reported in previous observations. The power
density spectrum (PDS) of the 5\,s time-resolution data shows a 
quasi-periodic oscillation (QPO) feature 
at 0.78\,mHz (=21.5\,min).  Assuming this periodicity represents the Keplerian
period at the  transition between the thin and advective disc regions, we
determine  the transition radius. 
We discuss the possible origins for the QPO feature in the context of the 
advection-dominated accretion flow model. 

We determine the colour of the large flares and find that the \sloani  
band 
flux per unit frequency interval is larger than that in the \sloang band.  
The colour is consistent with optically thin gas with a  
temperature of $\sim$8000\,K arising
from a region with an equivalent blackbody radius of at least 2\Rsun,  which
covers 3 percent of the accretion  disc's surface.  Our timing and spectral
analysis results support the idea that the rapid flares (i.e. the QPO feature)  
most likely arise from regions near the transition radius.

\end{abstract}

\begin{table*}
\caption{Log of \uc observations.}
\label{Table:log}
\begin{center}
\begin{tabular}{lcccccccl}
\hline
\noalign{\smallskip}
  Date    &  Run \#   & \# pts  & Exposure &  JD 2452520.5 + & 
Orbital phase & $g^{`}$  & $i^{`}$ & Comments \\ 
          &           &         & Time (s) &   start -- end~  & 
mid &  & & \\ 
\hline
\noalign{\smallskip}
  09/9/02 &  run 09(a) & 1200  & 5.0   &  26.96793 -- 27.05119  &  0.855  &
19.836 &  16.570    &  Poor seeing at end of night \\
  10/9/02 &  run 10(a) & 1266  & 5.0   &  27.87320 -- 27.94646  &  0.993  &
19.755 &  16.615    &  \\
  10/9/02 &  run 10(b) & 1152  & 5.0   &  27.97038 -- 28.03704  &  0.159  &
19.920 &  16.658    &  Good seeing \\
  12/9/02 &  run 12(a) & 1192  & 5.0   &  29.85483 -- 29.92379  &  0.293  &
19.545 &  16.406   &  \\
  12/9/02 &  run 12(b) & 1282  & 5.0   &  29.92406 -- 29.99824  &  0.310  &
19.641 &  16.467    &  \\
  12/9/02 &  run 12(c) &  848  & 5.0   &  29.99853 -- 30.04770  &  0.316  &
19.695  & 16.516    &  \\
  13/9/02 &  run 13(a) & 1082  & 5.0   &  30.83847 -- 30.90107  &  0.451  &
19.968 &  16.690   &  \\
  13/9/02 &  run 13(b) &  948  & 5.0   &  30.90135 -- 30.92339  &  0.460  &
19.879 &  16.656    &  \\
  14/9/02 &  run 14(a) & 8544  & 0.237 &  31.83812 -- 31.86152  &  0.602  &
-      &  16.768   &  \\
  14/9/02 &  run 14(b) & 2791  & 0.237 &  31.86162 -- 31.86902  &  0.605  &
-      &  16.669    &  Noisy data, not used in PDS \\
  14/9/02 &  run 14(c) & 9155  & 0.237 &  31.86936 -- 31.89443  &  0.607  &
-      &  16.649   &  '' \\
  14/9/02 &  run 14(d) & 7236  & 0.237 &  31.89942 -- 31.91924  &  0.611  &
-      &  16.754   &  \\
  14/9/02 &  run 14(e) & 8773  & 0.237 &  31.91932 -- 31.94334  &  0.615  &
-      &  16.708   &  \\
\noalign{\smallskip}
\hline
\end{tabular}
\end{center}
\end{table*}

\section{Introduction}
\label{Introduction}

Soft  X-ray transients  (SXTs)  are  a subclass  of  low-mass X-ray  binaries
(LMXBs) that  are characterized by episodic, dramatic X-ray  and optical 
outbursts, which usually last for several months and recur on a
timescale  of  decades. In the interim the  SXTs  are  in  a state  of
quiescence during which the optical  emission is dominated  by the luminosity 
of the faint  companion star \citep{Paradijs95}. In quiescence the  optical
lightcurves  exhibit the classical double-humped ellipsoidal modulation,
which is  due to the differing  aspects that the  tidally distorted secondary
star presents to the observer throughout its orbit \citep{Shahbaz03}.

In outburst the high/soft and the low/hard X-ray states are commonly seen. In
the high/soft state, X-ray emission is dominated by thermal emission from an 
accretion disc extending close to the last stable orbit around a black
hole.  In the low/hard state, the inner disc is believed to be truncated and
emission appears to arise from an extended corona.   Similar ideas are involved
for the advective models proposed 
for the quiescent state \citep{Narayan01}, but with the disc
truncated at larger radii. The states of SXTs are also classified by their
X-ray timing properties; the high/soft  state shows low level red noise, 
whereas
the low/hard state and very high states exhibit  band-limited noise but with a
low-frequency break at $\sim$0.02--30\,Hz, and sometimes  superposed QPOs (see
\citealt{Klis95} and \citealt{Wijnands99}).  

Although the spectral and timing properties during the quiescent state have not
been  well studied, one might expect similar properties, as the structure of the
flow is  believed to be similar to that in the low/hard state. The relatively
high quiescent X-ray luminosity of \target\ with respect to other SXTs 
indicates that there  is still some
continuing accretion. This is also supported by the short-term  variability
seen at all wavelengths in several quiescent photometric studies
of \target\ (\citealt{Pavlenko96}; Zurita, Casares \& Shahbaz 2003).
Sub-orbital variability with an  amplitude of 0.10--0.20 mag \citep{Wagner94}
which is partly due to a $\sim$6\,hr  quasi-periodic oscillation
(\citealt{Casares93}; \citealt{Pavlenko96})  is present in the optical. The
H$\alpha$ line profile also changes significantly on short timescales
\citep{Casares92a} and seems to be correlated  with the continuum
\citep{Hynes02}. Short timescale variations also appear in  the infrared
\citep{Sanwal96}. At X-ray energies, {\it ROSAT} (0.1--2.4\,keV)  saw changes
of a factor of 10 on timescales $<$0.5 day. More recent observations  with {\it
Chandra} (0.3-7\,keV) showed variability of a factor of 2 in a few ksec 
\citep{Kong02}. Finally \target\ is also a variable radio source on timescales
of  days, with typical variability in flux of 0.1-0.8 mJy \citep{Hjellming00}. 

The origin of the variability remains uncertain, the most likely explanations 
are magnetic reconnection  events in the disc or optical  emission from an 
advective region. The other possibilities such as  reprocessed X-ray variability
and flickering from the accretion stream impact point are less likely
\citep{Zurita03}. To
determine if just one type of variability dominates or if the variability is a 
combination of mechanisms on different timescales,  it is  therefore important
to perform a comparative study of the class.

In the optical, the faintness of the SXTs has usually limited the time
resolution  one can practically achieve. Furthermore, observations have
normally been  constrained to a single waveband. \uc with  4-m class
telescopes provides a unique opportunity not only to study 
photometric variability 
on the shortest timescales,  but also in more than one waveband 
simultaneously. Therefore we have embarked upon a programme to study the fast
multicolour  variability of quiescent SXTs with  \uc. Our main aims are to study the
multicolour variability  of the brighter SXTs and to study variability in the
fainter ones, faster than  previously possible. We present here high time
resolution multicolour observations  of one of the brightest SXTs, \target.

\section{Observations and Data Reduction}
\label{Observations}

Multicolour photometric observations of \target\ were taken with \uc on the
William Herschel Telescope atop La Palma during the period 
2002 Sep 9 to 13. \uc is an
ultra-fast, triple-beam CCD camera. The light is split into three wavelengths
colours (blue, green and red) by two dichroic beamsplitters and then passes through a
filter.  The detectors are three back-illuminated, thinned, Marconi
frame-transfer  1024$\times$1024 active area CCD chips with a pixel scale of
0.3\,arcsecs/pixel. The CCDs are  cooled using a three-stage peltier device and
water chiller, resulting in negligible  dark current (especially at these short
integration times). With the frame transfer mode, the dead-time is essentially
zero  (for further details see \citealt{Dhillon01}).

Our observations were taken using the Sloan 
\sloanu\,, \sloang and \sloani filters with
effective wavelengths of 3550\,\AA\,, 4750\,\AA\ and 7650\,\AA\ respectively.
For the first 5 nights (Sep 9 to 12), we used an exposure time of 5.0\,sec,
which was short enough to give reasonable counts in the \sloang and \sloani
bands.   Given the faintness of the object ($B$=20.6, $V$=18.4) and the short 
exposure times, few counts were obtained in the \sloanu\, band. 
On the last  night
(Sep 13) we decreased the exposure time to 0.237\,sec.  Only data in the
\sloani band were usable. From hereafter, we will refer to the data taken with
exposure times of 5.0\, sec and 0.237\,sec as the \slow and \fast data
respectively.

The weather conditions during the observing run were on the whole very good.
The night of Sep 9 suffered from severe seeing (2\arcsec) during the end of the
night.
No useful data could be obtained on Sep 11 due to cloud.
The night of Sep 10 was photometric, so we observed
the flux standard BD+35\,3659 \citep{Smith02} in order to determine the 
\sloanu, \sloang  and  \sloani band instrumental zeropoints, which were then
used to calibrate the data. On Sep 14, due to an earthing problem, the
electronic bias level  occasionally increased by  a factor of 2, however, this
only affected runs 14(b) and (c). A log of the
observations is given in Table\,\ref{Table:log}.

The pipeline reduction procedures were used to debias and  flat-field the data.
The same pipeline was also used to determine lightcurves for \target\ and
several comparison stars by extracting the counts using aperture photometry. As
there is a line-of-sight contaminating star 1.5\,arcsec North of \target, we
obtained the combined counts using an aperture  sufficient enough to encompass
both stars.  We determined lightcurves with different sized apertures, ranging
from  to 1.8 to 4.2\,arcsec. The optimal and most reliable results,  given the
contaminating star, were obtained using a large aperture of 3\,arcsec.
The count ratio of \target\ with respect to the local standard   
(3.6\,\arcsec\, S 75.8\,\arcsec\,E of \target)   was then determined by subtracting the count
ratio of the contaminating star with respect to the local standard determined
from images taken under good seeing conditions (0.7\,arsecs), 
from the combined count ratio
of \target\ (i.e. \target\ + line-of-sight star) with respect to the local
standard.  The local standard was  chosen to have a similar colour 
to the contaminating star.
The count ratio of the contaminating star to the local standard 
is 0.065  and 0.143 in  the \sloang and \sloani bands respectively.  
The magnitude of \target\ was then obtained using the observed magnitude of
the local standard, \sloang=16.90 and \sloani=15.12 ($<$0.01 mag uncertainty). 
The magnitude of the contaminating star was \sloang=19.87 and \sloani=17.23. 
As a check of the photometry and systematics in the reduction and extraction  
procedures, we also computed lightcurves of a comparison star  
(13.2\,\arcsec\,S, 17.1\,\arcsec\,E of \target).  
The magnitude of the comparison star was \sloang=18.90$\pm$0.03, 
\sloani=16.55$\pm$0.02.
The mean \sloang and \sloani band magnitudes of \target\ were 19.77 and 16.57
respectively.  We estimate the relative  photometric accuracy to be 3.0 and
0.9 percent  for the 'slow' \sloang and \sloani band respectively and 6.0
percent for the \fast \sloani band data.

By co-adding all the images from run 10(b) on Sep 10  taken under good seeing
conditions, we estimate  \sloanu=22.6$\pm$0.7 for \target.

\begin{figure}
\setcounter{figure}{0}
\label{Lcurves}
\psfig{angle=90,width=9.0cm,height=9.5cm,file=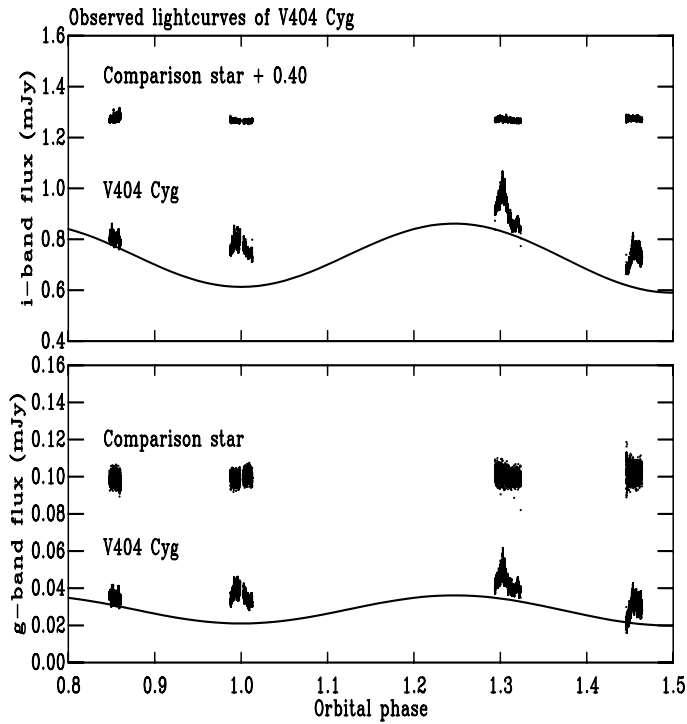}
\caption{ The observed \sloang and \sloani flux lightcurves 
of \target\ and the comparison star.
The solid line is the secondary star's ellipsoidal modulation (see text).}
\end{figure}

\begin{figure*}
\setcounter{figure}{1}
\label{Flares}
\vspace*{30mm}
\hspace*{-10mm}
\psfig{angle=0,width=8.5cm,height=11.0cm,file=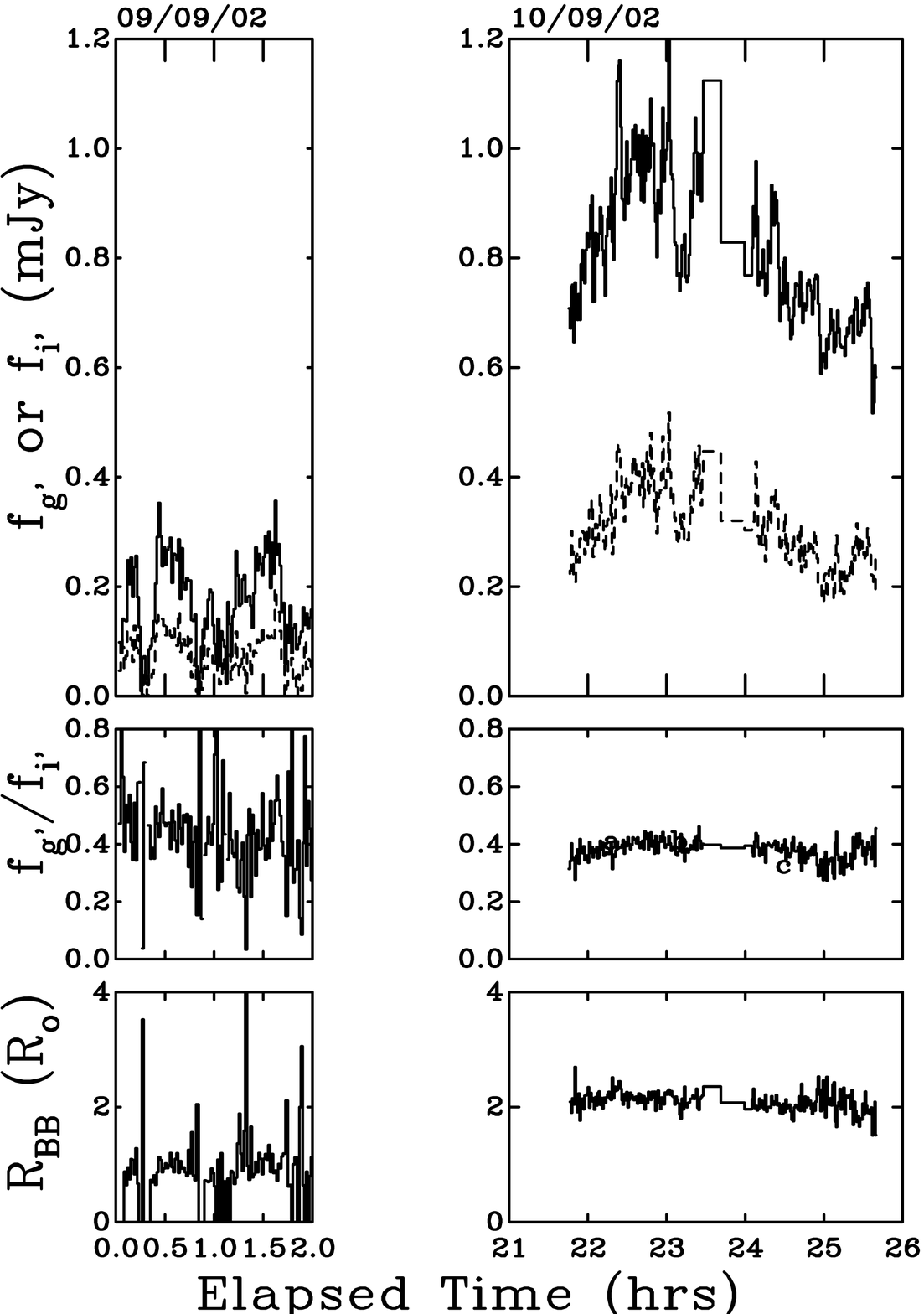}
\hspace*{10mm}
\psfig{angle=0,width=8.5cm,height=11.0cm,file=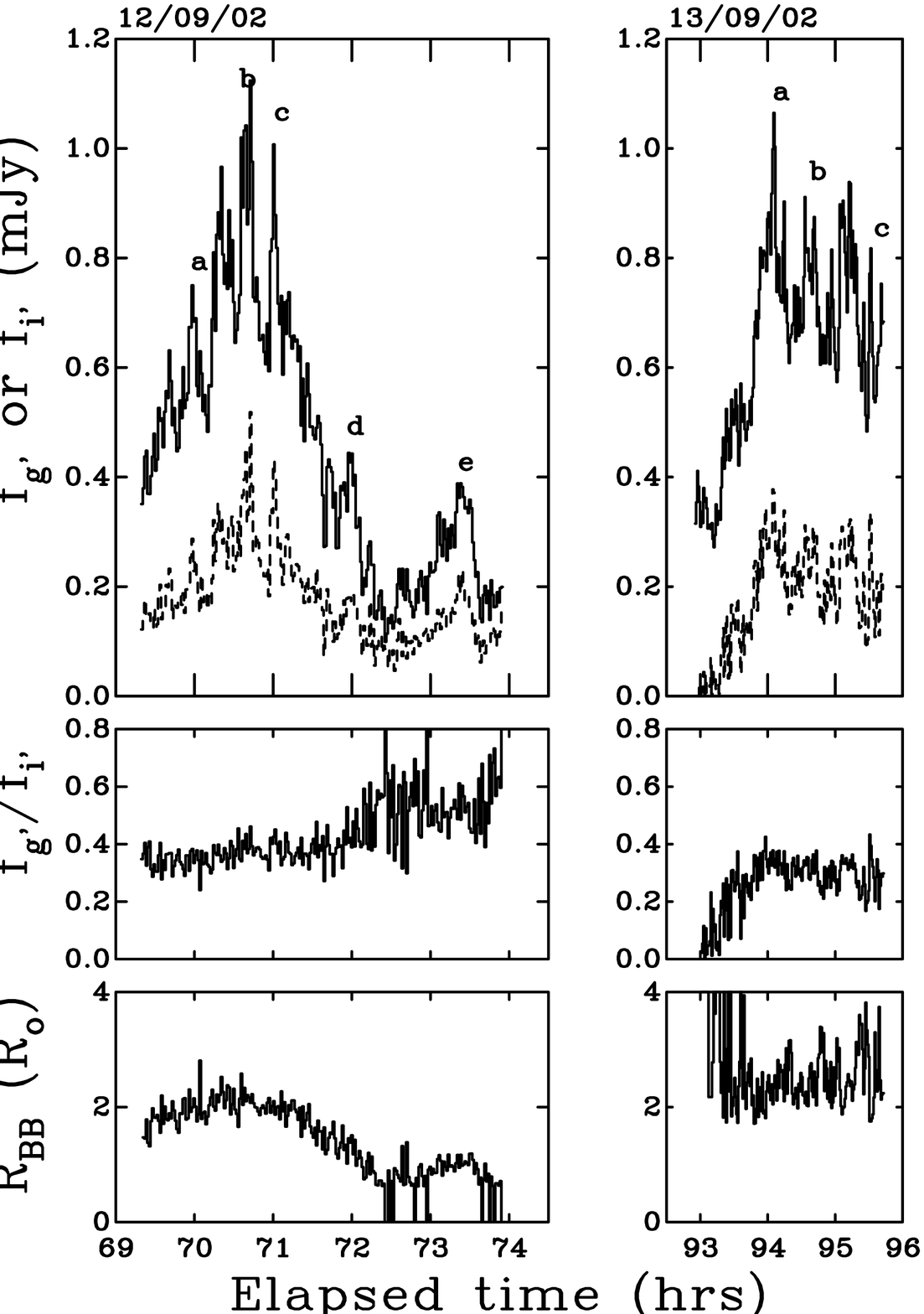}
\caption{
In the top panel we show the flare flux density lightcurve 
in the \sloang (dashed line) and \sloani (solid line) bands, 
obtained by subtracting the reddened
secondary star's ellipsoidal modulation from the observed lightcurves.
For clarity the flare lightcurves 
have been re-binned to a time resolution  1\,min. 
The uncertainties in the \sloang and \sloani lightcurves are
0.02\,mJy and 0.01\,mJy respectively. 
The letters mark the rapid flare events used to 
determine the flare properties in Table\,\ref{Table:flares}.
The middle panels show the flux density ratio $f_{g'}/f_{i'}$ and the bottom
panel shows the projected blackbody radius of the region producing the flares.}
\end{figure*}

\section{Background to \target}
\label{Background}

The X-ray transient GS\,2023+338 was discoverd during its ourburst in May 1989
by the All Sky Monitor aboard the Ginga satelite \citep{Makino89}. Its high 
X-ray luminonsity suggested that, like other X-ray transients, the system was a
close binary with an accreting compact object. The X-ray source 
was soon identifed
with the known variable star \target, which had been classified as a nova after
its 1938 outburst \citep{Marsden89}.

Spectroscopic observations taken when \target\ was in quiescence revealed the
companion to the X-ray source to be a G or early K-type star in an orbit with a
period of 6.473\,d,  which when combined with the secondary star's radial
velocity curve implies a mass function of 6.3$\pm$0.3\Msun \citep{Casares92b}. 
Later on, additional measurements  determined the secondary  star's rotational
broadening, and refined the spectral type of the secondary star and the
orbital parameters; the orbital period is $P_{\rm orb}$=6.4714\,d and
semi-amplitude of the secondary star's radial velocity curve is  
$K_{\rm 2}$=208.5\kms, giving a mass function of f(M)=6.08$\pm$0.08\Msun 
\citep{Casares94}. 
Since this mass function, which provides a firm lower limit to the compact
object's mass, exceeds the maximum mass of a neutron star ($\sim$3.3\Msun;
\citealt{Cook94}), the  compact object  in \target\ is 
almost certainly a black hole.

Optical photometry of \target\ in the $I$-band revealed the secondary star's
ellipsoidal variation \citep{Wagner92} but these were,
however, severely contaminated by short-term variability. 
Therefore, in an effort
to avoid these problems, \citet{Shahbaz94} observed the lightcurve of
\target\ in the $K$-band, where the flux from the K0IV star  is
substantially greater \citep{Shahbaz96}.   The analysis of the ellipisoidal
variations, which results from the tidal disortion of the secondary star gives
the binary inclination angle, combined with spectroscopic observations yields
the mass of the binary components.  They concluded that the contamination in
the infrared was small and deduced the black hole mass to be 12$\pm$2\,\Msun.

\citet{Casares93} estimated the interstellar reddening to \target, by comparing
its observed colours with those of a K0III star, after 
allowing for the accretion disc contribution to the observed flux. They
obtained $A_{\rm V} \sim 4$  and a lower limit of $A_{\rm V} \sim 3.6$. 
However, as one can see, the reddening determined in this way depends heavily
on the disc contamination in the optical, which the authors measure with large
uncertainites. A more accurate determination of the reddening was obtained by
\citet{Shahbaz94} who obtained the $K$-band lightcurve of \target\  in
quiescence and determined the binary inclination angle by modeling the
secondary star's ellipsoidal modulation. Assuming that the secondary star is a
stripped giant \citep{King93} they also determined limits to the reddening and
distance to \target, by  matching the model stripped giant luminosity 
to that of
the  secondary star as predicted by the ellipsoidal model. Assuming no
disc contamination in the $K$-band, a reasonable assumption  given that the
disc contamination measured in the $K$-band was found to be negligible
\citep{Shahbaz96}, and 10 percent in the $V$-band, they obtained    
2.2$<A_{\rm V}<$3.3 (90 percent confidence). 
Although the determination of the reddening through this method is model
dependant, the uncertainites  in the disc fraction and model dependant
parameters such as the binary masses and inclination angle are less in  the
infrared compared to in the optical.  Given the uncertainties in determining
the reddening using optical and infrared data, throughout this paper we   will
use a range of values for the reddening, 2.2$<A_{\rm V}<$4.0 and a central 
value of $A_{\rm V}=$2.8.

\section{The lightcurve of \target}
\label{Lightcurve}

The optical lightcurve of \target\ is dominated by the secondary star's
ellipsoidal modulation, which is due to the  observer seeing differing aspects
of the tidally  distorted secondary star.  This modulation gives rise to two
maxima and minima per orbital cycle.  However, superimposed on the ellipsoidal
modulation are the large 6\,hr  flares  seen previously \citep{Pavlenko96}
and also many rapid flares ($\sim$0.5\,hr),  similar to what is observed in the short
orbital period SXTs  (\citealt{Zurita03}; \citealt{Hynes03}). It is clear
that the secondary star's ellipsoidal modulation is not linked  to the 
variability, therefore if we want to determine the flux of the flares, its
contribution must be first removed  from the lightcurves. 

Figure~1 shows the observed data. 
Large amplitude flares which last a few hours or more, are observed only on  
Sep 10, 12 and 13  and are most likely related to the well known 6\,hr
flare/QPO  (\citealt{Casares93};  \citealt{Pavlenko96}; \citealt{Hynes02}).
On Sep 9 no large flare is seen.
In order to isolate the flux of the flare we assume that the light produced by
a flare is simply added to the quiescent spectrum, where the quiescent spectrum
is the light from  the secondary star's ellipsoidal modulation and the
non-variable accretion disc.  

We first de-redden the observed magnitudes
using the central value for  $A_{\rm V}$ of 2.8 (see section\,\ref{Background})
and the ratio  $A_{\rm V}/E(B-V)$=3.1 (\citealt{Seaton79}), giving \sloang and
\sloani extinction values of 3.34 and 1.79 mags respectively and then 
convert the Sloan AB magnitudes to flux density. 
We use the X-ray binary model described in \citet{Shahbaz03} with the binary
mass ratio and inclination angle given in \citet{Casares93} and
\citet{Shahbaz94} respectively, to determine the  ellipsoidal modulation. 
The model is not set up to compute Sloan magnitudes, so we first compute the 
Johnson-Cousins magnitudes and then convert to Sloan magnitudes using the
colour  transformations \citep{Fukugita96}.  
We shift the model ellipsoidal lightcurve so as 
to fit  the
lower-envelope of the observed lightcurve.  We use the data at phase 0.85 (Sep
9) to anchor the fit, since on  this night the large 6\,hr flare is not seen
and so the lower-envelope of these data most likely represents the light
from the secondary star and non-variable accretion disc. 
Figure~1 shows the data
and the scaled ellipsoidal model. Clearly the model fails to fit the \sloang
and \sloani data  near phase 0.0 (Sep 10). This suggests that the
accretion disc light is  variable, most likely due to a residual superhump
modulation as is observed in other quiescent SXTs (see \citealt{Leibowitz98};
\citealt{Zurita02}).  
This will obviously introduce uncertainties in the
subtraction procedure, however, we can obtain a good estimate for the 
flux of the
flares, since on Sep 12 and 13 the model fit  partially matches the end and
beginning of the large flares respectively.
Finally, the scaled ellipsoidal modulation is subtracted from the de-reddened
lightcurve yielding the colour of the flares (see Figure~2).  

One should note that the conversion from magnitudes to flux density is colour
dependant. In order to estimate the scale of this effect,  
we determine the fluxes obtained 
using spectra with different power-law indices  ($F_{\nu} \propto
\nu^{\alpha}$). The spectrum of the flares has a mean flux density ratio of
$f_{g'}/f_{i'} \sim$0.40  (see Figure~2) which corresponds to a 
spectrum with a power-law
index of $\sim$-2. The ellipsoidal model has a colour  (\sloang-\sloani)=1.67
which corresponds to a spectrum with a power-law index of -2.8.
Using these models, we estimate that the conversion from Sloan AB mags to flux
density units introduces an uncertainty in the colour $f_{g'}/f_{i'}$ 
of $\sim$10 percent.
It should also be noted that the residual superhump will introduce
an additional uncertainty in the colour of the flare. Since the 
superhump originates in the outer parts of the disc, it is most
likely to have a cool spectrum. 
Theoretical models for the behaviour of the quiescent accretion disc in 
\target\ indicates that the outer disc is optically thick and has a 
temperature of $\sim$2000\,K (Narayan, Barret \& McClintock 1997). 
If this also represents the colour of
the superhump, then assuming the superhump radiates as a blackbody,
we estimate that it would add an additional 
uncertainty of 2 percent in the determination of $f_{g'}/f_{i'}$.
Note that these  uncertainties are much less than that due to the 
reddening (see section\,\ref{FlareCharacteristics}).

\begin{figure}
\setcounter{figure}{2}
\vspace*{10mm}
\hspace*{15mm}
\label{Profile}
\psfig{angle=90,width=8.0cm,height=7.5cm,file=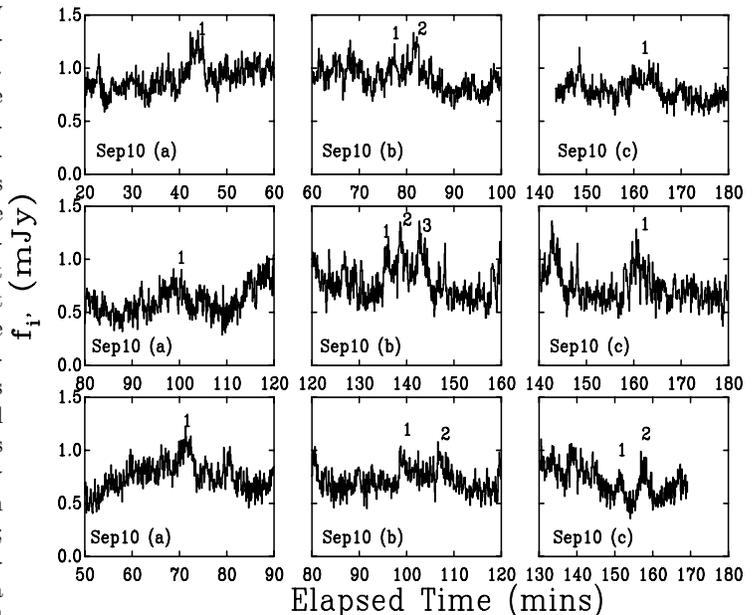}
\caption{Detailed plots of some individual distinct flares in the 
\sloani-band lightcurve. The labelling
refers to the events in Figure~2 and Table~2.}
\end{figure}

\subsection{The characteristics of the flares}
\label{FlareCharacteristics}

Although large amplitude flares are observed on more than one night,  it is
only on Sep 10 that  a large flare is observed throughout the event. 
The shape of the large flares seem to be symmetric 
with similar rise and decay times.  Superimposed on these large flares are many
rapid events which last typically $\sim$0.5\,hr. 
Furthermore, superimposed
on these flares, are still shorter term events on  timescales of mins which 
show unresolved peaks (see Figure~3). 
This makes it difficult to determine the exact
shape of the flare events; 
there is an indication that some of the flares on the shortest timescales
have a burst-like profile [e.g. see flare (b) on Sep 13]. 
However, in general all the flare
events appear to have a linear rise and decay. In Table\,\ref{Table:flares} we
estimate some of the characteristics of  the individual distinct flares which
are shown in  Figure~3. 
The flares typically have a rise time of
2\,mins and a duration of 6\,mins. 

At first sight, once clearly notices that the monochromatic flux in \sloani
is larger than that it \sloang,  at the peak of the large flares,  i.e the
flare at its peak has a  relatively red spectrum. In general the colour flux
ratio $f_{g'}/f_{i'}$ of these flares increases, becoming more blue at the 
peak. After the peak is reached, the flux and colour decreases
becoming more red.  The exception is the large flare on Sep 13; the colour
seems to increase continuously as the flare proceeds. 
The most  prominent and distinct rapid flares are marked
in  Figure~2. 
The mean flux ratio $f_{g'}/f_{i'}$ of the large flares
is $\sim$0.4. If we use the extreme values for the reddening $A_{\rm V}$=4.0
and $A_{\rm V}$=2.2,  then the flux ratio $f_{g'}/f_{i'}$ changes
by a factor of 0.53 and 1.36 respectively. 

To test if the flaring in the \sloang and \sloani bands are simultaneous we 
calculated the cross-correlation function. The cross-correlation function was
symmetric and no time lag was found,  so we conclude that the flaring in the
two bands is simultaneous (see Figure~4).

\begin{figure}
\setcounter{figure}{3}
\label{XCCF}
\vspace{-10mm}
\hspace{10mm}
\psfig{angle=90,width=10cm,file=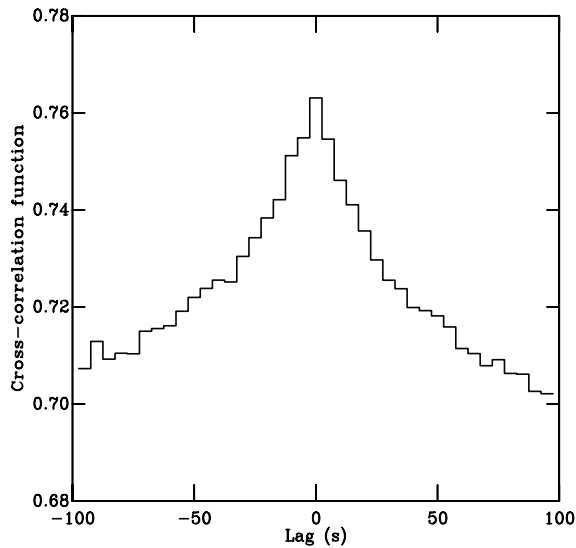}
\caption{Cross-correlation function between the \sloang and \sloani-band
lightcurves. A postive lag would indicate that the \sloang variations lag
behind those in  \sloani.}
\end{figure}

\begin{table}
\caption{Properties of the \target\ flare events. 
The rise and decay times are obtained from linear fits to the 
\sloani band flare profile. }
\label{Table:flares}
\begin{center}
\begin{tabular}{lcccc}
\hline
\noalign{\smallskip}
Date & Flare   & Rise & Decay &  PDS power-law index$^{*}$  \\
     & No.     & (s)  &  (s)  & \sloang ~~~~~~~ \sloani   \\
\hline
\noalign{\smallskip}
Sep 10 & a$^1$ &   240  &  110  & -1.23 ~~~~~ -1.56 \\
       & b$^1$ &   104  &  100   \\
       & b$^2$ &    57  &   58   \\
       & c$^1$ &    65  &   55   \\
Sep 12 & a$^1$ &   322  &  242  & -0.96 ~~~~~ -1.32 \\
       & b$^1$ &    62  &   54   \\
       & b$^2$ &    88  &   44   \\
       & b$^3$ &    62  &  187   \\
       & c$^1$ &   175  &  241   \\
Sep 13 & a$^1$ &   170  &  136  & -1.17 ~~~~~ -1.49 \\
       & b$^1$ &    20  &   96   \\
       & b$^2$ &    22  &   78   \\
       & c$^1$ &    90  &  130   \\
       & c$^2$ &   140  &  140   \\
\noalign{\smallskip}
\hline
\end{tabular}
\end{center}
\noindent
\noindent
$^*$Power-law index 1-$\sigma$ errors of 0.03. 
\end{table}

\section{The power density spectrum}
\label{PowerSpectrum}

To compute the PDS of the \slow data, we use the same flare lightcurves as 
described in section\,\ref{Lightcurve}, but since we are only interesting in
sub-orbital variability, and in order to preserve the  flux in each band, we
detrend the data for each night using the nightly mean  and add on the mean 
flux level of all the data.  Although the sampling of \uc is perfectly uniform,
we use  the method of Lomb-Scargle to compute the  periodograms
\citep{Press92}.  The Lomb-Scargle method was chosen so that we could combine
consecutive runs and confidently compute the periodogram.  Also, in order to
allow a direct comparison with X-ray PDS, we use the same normalisation method
as is commonly used in  X-ray astronomy, where the power is normalised to the
fractional root mean amplitude squared per hertz.  The advantage of this kind
of normalisation  is that fractional rms amplitudes can be directly estimated
from the level of the PDS \citep{Klis95}.   

To compute the PDS we use the constraints imposed by the Nyquist frequency  and
the typical duration of each observation. We then average the PDS for each
observing sequence. We bin and fit the PDS in logarithmic space
\citep{Papadakis93} and the errors in each bin are determined from the standard
deviation of the points within each bin.  The white noise level was subtracted
by fitting the highest frequencies with a white noise plus red noise model.

\subsection{The PDS of the ``slow'' data}
\label{SlowPDS}

To determine the PDS for the \slow data we only use the data with the longest
baseline i.e. during the nights of Sep 10 and 12. This allows us to 
extend the PDS to the lowest possible frequencies. The derived PDS is shown in
Figure~5.  The shape of the \sloang and \sloani band  PDS are
well  described by a power-law with slopes of $-$1.24$\pm0.02$ and  
$-$1.37$\pm$0.02 respectively  (1-$\sigma$ errors). 
On each individual night, the slope of the power-law changes 
(see Table\,\ref{Table:flares}).  
To examine the systematic effects in our analysis, we also
computed the  PDS of the comparison star.
The PDS was found to be flat, exactly  as expected
for white noise data. Furthermore, the auto-correlation function of the 
comparison star is a delta function, implying that the data are not 
correlated. 
A QPO feature with a centroid frequency at 0.78\,mHz (= 21.5\,mins) is present,
not surprising given that one can clearly see variability on timescales of 
0.5-1.0\,hrs in the lightcurves. The fractional root mean  squared amplitude 
is 3 and 1.3 percent for the \sloang and \sloani band QPOs and was computed by 
integrating the Gaussian function and taking the square root. 

\subsection{The PDS of the ``fast'' data}
\label{FastPDS}

We used run14(a), run14(d) and run14(e) to compute the PDS. The other runs
were not included because of the high noise level of the data,  the result of an
increase in the electronic bias level  (see section\,\ref{Observations}). 
The computed \fast PDS is shown in Figure~5.  
The general shape of the PDS has a power-law shape   with a slope of
$-$1.16$\pm$0.04 (1-$\sigma$ errors).   As a check we also computed the  PDS of
the comparison star.  Again the derived PDS was flat, which is exactly  as expected
for white noise data (see Figure~5).

\begin{figure}
\setcounter{figure}{4}
\label{PDS}
\vspace{-10mm}
\hspace{10mm}
\psfig{angle=90,width=10cm,file=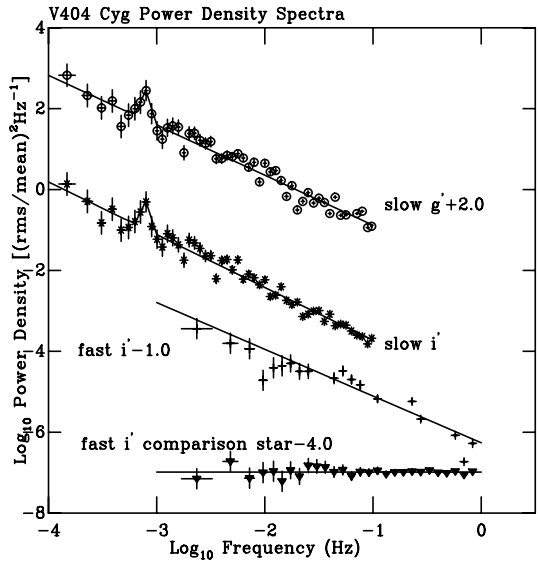}
\caption{The power density spectrum of \target. 
From top to bottom: \target\ \sloang and \sloani\,\slow\,PDS, 
\target\ \sloani\,\fast\,PDS and the comparison star PDS.}
\end{figure}

\section{Discussion}
\label{Discussion}

\subsection{The slope of the PDS}
\label{SlopePDS}

The outer regions of the accretion discs in quiescent SXTs and dwarf  novae
(DNe) are assumed to be very similar. Therefore if the optical variability 
observed in the quiescent SXTs originates from these outer regions,  then they
should have similar properties to those observed in quiescent DNe.
\citet{Bruch92} has performed a study of  the rapid variability in cataclysmic
variables. The slope of the  PDS in quiescent DNe (--1.6 to --2.6) seems to be
in general steeper than those observed in quiescent SXTs [-1.0 to
-1.52; \citet{Zurita03} and \citet{Hynes03} ]. 
This suggests that there is either more low-frequency variability 
in the lightcurves of quiescent DNe compared to quiescent SXTs, or that
there is more high-frequency variability in the SXTs. 
Note that the former could 
be attributed to intrinsic variability associated with the 
stream/disc impact region or the white dwarf (e.g. \citealt{Warner95}).

On average the slope of the \sloani PDS in \target\ is steeper than the 
\sloang slope, suggesting that there is more low-frequency disc 
variability in \sloani compared to \sloang.  This difference
could be attributed to more variability from outer regions of the disc, such as
the stream/disc impact region or more high frequency variability
in \sloang from the inner disk. Thus one expects the slope of the PDS to be
dependent on the observed waveband.

\citet{Hynes03} determined the PDS of the quiescent SXT A0620--00 and found a
possible break in the PDS at 0.95\,mHz, a feature which is also seen in the 
low/hard state of SXTs (see \citealt{Wijnands99} and references therein).
Although the origin of the break frequency is uncertain, it could be related
to  the size of the inner advective region; a low break frequency in quiescence
would then  be expected, as the advective region should be larger.    If the
break frequency is associated with a dynamical timescale, then it should vary
as $R^{1.5}$, where $R$ is the radius of the advective region  
(see section\,\ref{QPO-ADAF}).  
Furthermore, as noted by \citet{Hynes03} the break frequency most  
probably scales with some characteristic length scale, which could  depend on
other parameters, as well as the inner disc radius.  

\subsection{The QPO feature and the ADAF model}
\label{QPO-ADAF}

\citet{Narayan96} and \citet{Narayan97} showed that 
the observations of 
quiescent BH SXTs can be explained by a two-component accretion flow model. The
geometry of the flow consists of an inner hot advection-dominated accretion
flow  (ADAF) that extends from the black hole horizon to a transition radius 
\rtr=$10^{4-5}$ \rsch\ (\rsch= $GM/c^{2}$ is the Schwarzschild radius for
a black hole with mass $M$)  and a thin accretion disk that extends from 
\rtr\ to an outer radius \rout\, i.e. the radius of the accretion disc.  A key feature
of the ADAF model of quiescent BH SXTs is its low radiative efficiency. 
The bulk of the viscously dissipated energy  is stored in the gas and is
advected with the flow into the black hole. This naturally explains the
unusually low luminosity of BH SXTs, since in NS SXTs,  all the advected energy
is radiated from the neutron star surface, resulting in a  much higher overall
radiative efficiency \citep{Narayan95}.

An ADAF has turbulent gas at all radii, with a variety of timescales, ranging
from a slow timescale at the transition radius down to nearly the free-fall
time close to the black hole.  In principle, interactions between the hot inner
ADAF and the cool, outer thin disk,  at or near the transition radius,  can be
a source of optical variability, due to synchrotron emission by the hot
electrons in the ADAF. For an ADAF the variability could be  quasi-periodic and
would have a characteristic timescale given  by a multiple of the Keplerian 
rotation period at 
\rtr\,  $t_K = 2\pi R/ v_K = 2\pi(GM/R^{3})^{-1/2} \sim 
(M/10\,M_{\odot})(r/100)^{1.5}$\,s.  
where $R$ is the absolute radius and $r$ is in units of \rsch. One also expects
slower variations, since the mass supply to the ADAF originates  at 
\rtr \citep{Narayan97}. It should be noted that it is difficult 
to produce such rapid variations using the thin disc models. \citet{Narayan97} 
used the maximum velocity observed in the H$\alpha$ emission line arising from
the   quiescent thin accretion disc to estimate \rtr. However, they note that 
measuring the maximum velocity is difficult, and so 
their  estimate of 10$^{4.4}$\rsch\ for \rtr\ should 
only be considered as an upper
limit. If we assume that the 0.78\,mHz QPO feature observed in the  PDS (see
section\,\ref{PowerSpectrum}) is the dynamical timescale at the transition 
radius, then \rtr\ lies at 10$^{4.0}$\rsch, which is consistent with
that estimated by \citet{Narayan97}. We can write \rtr\ in terms  of the
accretion disc radius.  For the binary parameters of \target\ the outer
radius  of the  disc  lies at \rout=$2.3\times10^5$\rsch\ or 12\Rsun,
and so \rtr\ lies at 0.04\Rdisc or 0.5\Rsun.

\subsection{Comparison of the \target\ QPO features with 4U\,1626-67}
\label{QPO-Other}

It is interesting to note the similar timescales of the QPOs we have
discovered in \target\, with those found in various X-ray/UV/optical
observations of the double-degenerate LMXB, 4U\,1626-67. 
\citet{Chakrabarty01} used fast CCD photometry to 
reveal the presence of 3 oscillations: the 130\,mHz pulsation 
(i.e. the well-known 7.66\,s
X-ray/optical pulsation of the neutron star), a 48\,mHz ($\sim$20\,s) QPO (a
broad oscillation seen before in X-ray, optical and UV wavebands) and a
1\,mHz QPO.  This latter oscillation had not been seen before, is not
present in any X-ray observations, is quasi-sinusoidal in appearance, and
is dramatically stronger in far UV HST data.  \citet{Chakrabarty01} concluded
that the first 2 oscillations were clearly X-ray driven and the optical/UV
analogues were due to reprocessing in the disc/companion star.

However, the 1\,mHz QPO origin is extremely puzzling.   The timescale is
actually the same as the pronounced X-ray and optical flaring seen from
4U1626-67 during the 1970s and 80s, but this flaring behaviour has
vanished since the torque reversal that occurred in 1990.  With no X-ray
oscillation to drive this process, \citet{Chakrabarty01} suggested that the
1\,mHz QPO is due to ``warping'' of the inner accretion disc and is of an
appropriate timescale for the radius of the inner disc that is set by the
magnetospheric corotation radius.  In this way, the modulation would be
expected to be much greater in the far UV (inner disc) than the optical
(in the outer, shielded regions).
It is possible that, in spite of the fact that \target\, almost certainly
hosts a $\sim$12~M$_\odot$ black hole, this analogy can be applied from
4U\,1626-67.  The difference is in the origin of the truncated inner disc,
which for \target\, in X-ray quiescence is expected to be in an ADAF phase
\citep{Narayan97}.

\subsection{The emission mechanism of the flares}
\label{FlareEmission}

As one can see from Figure~2, on most of the nights there 
are numerous rapid flares ($<$0.5\,hr) superimposed on the single large
event ($\sim$few hrs). Given the uncertainties in the data, 
it is difficult to determine the shape and hence the physical  parameters of 
these rapid events.
We can however comment on the general physical properties of the flares
using the large events  ($\sim$few hrs).
Using the flare flux density lightcurves we determine the flux density
ratio $f_{g'}/f_{i'}$ and the blackbody temperature $T_{\rm BB}$ and 
equivalent radius $R_{\rm BB}$. 
To determine the  colour temperature corresponding to a given flux density ratio, 
we integrate blackbody functions with the CCD and Sloan filter response
functions  and then determine the $f_{g'}/f_{i'}$ flux density  ratio. Given
this blackbody temperature we can then determine the  corresponding
radius of the region that produces the observed dereddened flux at a 
distance of 3.5\,kpc \citep{Casares93}. 
Also, using the flux density ratio, the power-law index
$\alpha$ of the spectrum
($F_{\nu} \propto \nu^{\alpha}$) can be obtained.
One can easily show that the power-law index is given by 
$\alpha = \Delta \log F_\nu / \Delta \log \nu = 4.83\log(f_{g'}/f_{i'})$.

For Sep 9, the flux density ratio $f_{g'}/f_{i'}$ remains constant at 0.5; 
$\alpha$ and $T_{\rm BB}$ have constant values of 
$\sim$-2 and 5100\,K respectively.
For the single large flare events on Sep 10 and 13, 
$f_{g'}/f_{i'}$ is correlated with the flux of the flare; 
$f_{g'}/f_{i'}$ peaks at $\sim$0.3; 
$\alpha$ and $T_{\rm BB}$ peaks at $\sim$-2.5 and 4100\,K respectively.
On Sep 12 one can see that the blackbody radius
and flare flux are correlated i.e. as the flare flux increases and
decreases so does the equivalent blackbody radius. The blackbody radius
increases from $\sim$1\Rsun to $\sim$2\Rsun. 
The same is seen during the decay of
the large flare on Sep 10, but to  a lesser extent.
However, it should be noted that the uncertainties in the reddening 
(see section\,\ref{Background}), 
which typically correspond to $\pm$700\,K in the $T_{\rm BB}$ 
estimates, are such that we 
cannot comment on any colour or temperature variation during the flares.

The peak flux produced by the large flare arises from a region that 
has a projected blackbody radius of 2\Rsun\, and the rapid flares most 
likely from smaller regions.
In other words, the emitting area at the peak of the large flare covers 
at least 3 percent of the total disc emitting area.
Note that given the data presented here, we cannot state where 
in the disc these large flares arise. 
However, it is most likely that the large flares
originate in the outer regions of the disc (see section\,\ref{FlareWhere}).

During the large flares observed in \target\ the power-law index of the 
spectral energy distribution is $\sim$--2.0 (a range of --1.5 to --2.5, 
depending on the flare).
It is difficult to explain this in terms of optically thin synchrotron
emission, unless the electrons follow a much steeper power-law energy
distribution ($N(E)dE \propto E^{p}dE$). For solar and stellar flares the
power-law index of the electron energy distribution $p$ is $\sim$--2
\citep{Crosby93}, which
corresponds to a frequency spectrum with a power-law index of $\alpha$=--0.5
($F_{\nu} \propto \nu^{(1+p)/2}$).  The frequency power-law index oberved in
\target\ $\alpha$ of $\sim$--2.0 implies a much steeper index  for the  electron
energy distribution of $p$=--5.
However, given
the uncertainites in the reddening and therefore in the flux density ratio
$f_{g'}/f_{i'}$ (see section\,\ref{Background}),  we find a 
range of values for $p$ of --3.6 to --7.5.
It could be that the electrons are thermal and thus do not follow a power-law
energy distribution. Also, given that the  geometry and physics around a black
hole is presumably more extreme, such a steep index for the electron energy
distribution is not completely implausible.

The simple model for the flares  of blackbody radiation from  a heated region
of the disc's photosphere (i.e. photospheric in origin) is unlikely.
A more likely model for a thermal flare is emission from an optically thin
layer of recombining hydrogen, which is essentially the mechanism generally
accepted for solar flares.  We therefore  determine the continuum emission
spectrum of an LTE slab of hydrogen and then calculate the expected flux
density ratio  $f_{g'}/f_{i'}$ in the \sloang and \sloani bands using the same
method  as the blackbody case.  We find that a continuum slab temperature of
8000\,K with a baryon column density of  10$^{22}$\,cm$^{-2}$ (taken from 
\citealt{Welsh93}) gives $f_{g'}/f_{i'}$=0.30; propagating the  uncertainty in
the reddening (see section\,\ref{Background}), we find the range  for the slab
temperature to be 5500\,K and 9100\,K. It should be noted that the hydrogen
slab model does not include the contribution from Balmer  emission lines. 
Since the
\sloang filter includes  H$\beta$, our determination of the flux density ratio
$f_{g'}/f_{i'}$  is biased. However, we can estimate the amount of bias given
the H$\beta$ flux and equivalent width, 0.20\,mJy and 10\AA\ respectively 
\citep{Casares93}. Since the \sloang filter has a width of $\sim$2000\AA\, the
contribution of the H$\beta$ emission line to the \sloang flux density in
\target\, is only 1 percent, in constrast to that in AE\,Aqr \citep{Welsh93}.

Since we only have observations in two wavebands, we cannot accurately determine the physical
parameters of the flares.  However, it should be noted that the spectrum of the
flare seems to be similar to that observed in AE\,Aqr
\citep{Welsh93}.  Using the colour information for the large flare 
events, we  can say that the large flares arise from a region which  extends  a
projected area which has an equivalent $R_{\rm BB}$ of $\sim$2\Rsun.  One
should regard the equivalent $R_{\rm BB}$ estimate  as a  lower limit,
because the emission mechanism is unlikely to be blackbody. 
In general the rapid
flares seem to  arise from optically thin gas which has a peak
temperature of $\sim$8000\,K in the continuum. 
However, to obtain a more accurate estimate of the physical conditions of the
flares, one must resolve the Balmer jump and Paschen continuum.

Spectroscopic observations suggest that the whole disc participates  in the
large flares, since the whole H$\alpha$ emission line  profile changes
\citep{Hynes02}. This seems to contradict our earlier results,   which suggest
that the large flares arise from regions that cover only  3 percent of the
disc's total  area. However, this most likely indicates that the emission is
optically thin, or that the emission area  is small. Also, given the low
resolution of the spectroscopy, we can only really say that material from all
around the disc is contributing, given that the  line profile changes observed
extend across the whole line profile and  appear double-peaked. It is not
possible  to exclude  the flares from arising from only the inner regions of
the disc.

\subsection{Where are the flares produced?}
\label{FlareWhere}

\citet{Zurita03} have presented arguments that suggest the most likely 
mechanism for the flares is  by  local  magnetic
reconnection events  in or above the accretion disc.  It is  believed that a
dynamo mechanism, driven by the strong shear produced by differential rotation,
operates in accretion discs (\citealt{Tout92}). Regions of oppositely
directed magnetic fields develop within the disc or between the disc 
and corona and reconnect explosively, resulting in a flare 
\citep{Haswell92}. 
The duration of  the flare is directly related to the shearing  timescale.
As suggested by \cite{Zurita03}, assuming that the  range of flare
durations observed in the SXTs reflect their location  in the accretion disc,
then the  flares with  the  longest duration represent the shearing  timescale
in the outer parts of  the disc.  
Thus the maximum flare duration should be
limited by the outer disc radius and \target\ should show short-term 
variability, similar to what is observed in A0620--00. 
The very rapid flares ($\sim$\,mins) observed in \target\ (see Figure~3) have 
a similar timescale to those 
observed in A0620--00 \citep{Hynes03} and seems to support the scenario that 
the large  flares ($\sim$ few hrs) are produced in regions
further out in the disc than the rapid  ($\sim$0.5\,hr) 
and very rapid ($\sim$\,mins) flares.

Using the colour information about the large  events ($\sim$\,hrs), we can say
that the large flares are produced from a region which extends an equivalent
blackbody radius of $\sim$2\Rsun. We cannot determine the emitting area of the
rapid flares ($\sim$0.5\,hr) as the result may be biased due to the large 
flares. However, if our interpretation of the 21.5\,min QPO feature is correct,
then these rapid flares are most likely produced near the transition radius of
the ADAF, in the inner parts of the disc.  We cannot determine the exact
mechanism for the flare production. The flares could arise in optically thin
gas with a temperature of $\sim$8000\,K, most likely in a corona above the
accretion disc, or  from optically thin synchrotron emission. 

\citet{Nayakshin01} have argued that an ADAF  model is not required to
reproduce the observed  X--ray luminosities of quiescent SXTs. They propose a
model which consists of a standard thin disc with a hot corona,  powered by
magnetic flares. Although this scenario seems appealing,  it is difficult to
see how the thin disc model could produce rapid variations.   However, it
should be noted that the explosive magnetic reconnection model described in
\citet{Haswell92} could produce rapid variations from a  thin disc, similar to
an ADAF. Our observations suggest that for the rapid flares ($\sim$0.5\,hr)  
at least, they
are  produced  near the transition radius.  However, it is only by obtaining
precise  velocity information about the flares, can we  begin to place
constraints on the exact  location of the flares  in the accretion disc.

%
%

%
%
\section*{Acknowledgments}
TS acknowledges support from the Spanish Ministry of Science and Technology 
under project AYA\,2002\,03570.
RIH is currently supported by NASA through Hubble Fellowship grant 
\#HF-01150.01-A\ awarded by the Space Telescope Science Institute, which 
is operated by the Association of Universities for Research in Astronomy, 
Inc., for NASA, under contract NAS 5-26555.
VSD and TRM acknowledge the support of PPARC through the 
grant PPA/G/S/1998/00534.
Based on observations made with  the William Herschel Telescope  
operated on the island of La Palma by the Issac Newton Group
in the Spanish Observatorio del
Roque de los Muchachos of the Instituto de Astrof\'\i{}sica de Canarias.

%
%

%
\end{document}